# Interplay of Magnetism, Band Gap Tuning, Optical, and Thermoelectric Responses in Fe-Doped YMnO$_3$: Insights from First-Principles Calculations

Kazi Mazba Kamal and Alamgir Kabir *

Department of Physics, University of Dhaka, Dhaka – 1000, Bangladesh.

Email: *alamgir.kabir@du.ac.bd*

## Abstract

The hexagonal multiferroic oxide YMnO$_3$ has demonstrated applications in various fields and is widely researched due to its interesting properties. Since Mn(3d)–O(2p) interactions predominate close to the Fermi level, doping in the B-site (Mn) with Fe provides a way to modulate the band gap and magnetic order of YMnO$_3$. The need for a lead-free ferroelectric material with a narrow band gap is crucial for absorbing a wide range of the solar spectrum. Density functional theory calculations were carried out using GGA and meta-GGA (for an accurate description of the band gap) exchange correlation functional for the Fe-doped YMnO$_3$ multiferroics. Various magnetic configurations were analyzed, finding collinear G-type AFM as the least energy state. The hexagonal lattice is retained after Fe doping with slight distortions and a change in lattice constants. Fe doping reduces spin frustration and induces magnetization, while reducing the band gap from 1.88 eV for pure to 1.19 eV for a 25% doping concentration. Additionally, Fe doping exhibits an enhanced dielectric response, characterized by an increase in the static dielectric constant and the presence of strong absorption peaks in the visible and UV energy ranges. Thermoelectric studies illustrate enhanced conductivity due to increased charge carriers induced by doping. In summary, first-principles predictions of structural, electronic, optical, and transport behavior in Fe-doped YMnO$_3$ provide a foundation for tailoring this oxide in photovoltaic, thermoelectric, and optoelectronics applications.

# Introduction

Rare earth transition metal oxides (ABO$_3$) are typically known for their multiferroic properties, which are a special class of materials that exhibit two or more ferroic orders (ferroelectricity, ferromagneticity, ferroelasticity, antiferromagnetism, etc.) simultaneously within the same phase[1]. This type of material often exhibits a piezoelectric response (in the case of simultaneous ferroelectricity and ferroelasticity) or a magnetoelectric response (in the case of simultaneous ferroelectricity and ferromagnetism)[2]. Materials that have atoms with half-filled d shells usually contribute to ferromagnetism, and atoms with empty d shells contribute to ferroelectricity[3]. For example, the rare earth oxides BiFeO$_3$ have both of these features, which is why the multiferroic properties are seen in these materials[4]. Magnetoelectric coupling is of special interest to researchers as it paves the way to potential applications in memory devices. Rare-earth manganese oxides (RMnO$_3$) belong to the class of multiferroic materials with desirable physical characteristics resulting from the interaction between magnetic and ferroelectric order parameters. YMnO$_3$ is a widely studied rare-earth manganese oxide[5–7], which is predicted to adopt cubic, orthorhombic, and hexagonal structures[8–11]. These different forms appear depending on the external conditions: the hexagonal phase is stable at ambient pressure and changes with temperature[12], while the orthorhombic phase is usually obtained under high pressure[8], and the cubic phase is mostly discussed in theory[11]. YMnO$_3$ is characterized by a hexagonal structure at room temperature, with a space group P6$_3$cm[13]. The h-YMnO$_3$ exhibits a ferroelectric transformation below T$_C$ ~ 930K and an anti-ferromagnetic order below T$_N$ ~ 70K[14]. Another transition was observed, resulting in a centrosymmetric P6$_3$/mmc space group at 1258 ± 14K[15]. However, it shows improper ferroelectricity, resulting from geometric frustration[16]. Polarization P$_s$ = 6 µC/cm$^2$ results from the tilting of the Mn-centered oxygen octahedra and buckling of the Y-O planes, with no significant off-centering of the Mn cations as would be characteristic of a perovskite ferro-electric[16,17]. As such, YMnO$_3$ has been referred to as a "geometric ferroelectric"[18]. Due to the multi-ferroic nature, YMnO$_3$ has applications in magnetic random-access memory[19–21], colossal magnetoresistance (CMR)[22], photocatalytic materials[23], magnetic refrigeration technology[24], electrical field sensors[25], and photovoltaic materials[26]. Although there are numerous studies on YMnO3, there are very few studies on doped YMnO3[27]. Doping on YMnO$_3$ can potentially open up a lot of new possibilities. It was found that YMnO$_3$ exhibits a hexagonal-to-rhombohedral phase transition upon substitution of Ti for Mn[28]. In a recent study, it has been shown that the band gap of YMnO$_3$ can be lowered by Os doping on the B site [29].

As Mn-O bonding makes the most significant contribution near the Fermi energy, substituting Fe on the Mn (B site) site can enable us to tune the band gap of YMnO$_3$, which could lead to possible applications in photovoltaic devices. To the best of our knowledge, there is no computational study on the band tuning of $YMnO_3$ via Fe doping at the B site. By introducing Fe at the Mn site, it is possible to tune the electronic structure since the d-orbital occupancy and charge states are modified. This can help to reduce the band gap, adjust carrier concentration, and even improve magnetoelectric coupling[30]. It is anticipated that the active layer material shall have a small band gap to absorb a broad range of solar radiation because photovoltaic devices use the solar energy to generate photocurrent. Because of their robust inversion symmetry breaking, ferroelectric materials enable the desired separation of photo-generated charge carriers and permit voltages above the bandgap, potentially enabling efficiency above and beyond what is achievable in a traditional p-n junction solar cell[31]. A significant bandgap acts as a limiting factor for ferroelectric photovoltaics, and the reduction of bandgap is crucial for maximization of photocurrent[32]. In addition, both experimental and theoretical studies have reported that Fe substitution leads to mixed valence states ($Mn^{3+}/Mn^{4+}$ and $Fe^{3+}/Fe^{4+}$), which strongly affect electronic transport, spin reorientation, and dielectric behavior[33]. From an application point of view, such band structure tuning can increase the potential of $YMnO_3$-based materials for solar energy harvesting, non-volatile memory, and other multifunctional oxide devices. Additionally, a common concern of perovskite photovoltaic cells is the inclusion of lead as an element, which raises environmental and health issues[34]. Rare-earth oxides are lead-free perovskite options, which are not prone to further oxidation. Therefore, a first-principles investigation of Fe doping at the B-site offers a promising route to understand and control the band structure, while also shedding light on the intricate interplay between electronic, magnetic, optical, and transport properties.

## Computational Methodology

The Vienna ab initio simulation package (VASP) has been used to perform the density functional theory-based calculations, implementing the projector augmented-wave method (PAW)[35–37]. Structural optimizations are performed using the generalized gradient approximation (GGA) for the exchange-correlation functional, into the Perdew-Burke-Ernzerhof (PBE)[38]. For an accurate representation of the band gap, we have extended the calculation to meta-generalized gradient approximation (meta-GGA) employing Modified Becke-Johnson potential (mBJ)[39,40]. The Y (4s, 4p, 4d, 5s), Mn (3p, 3d,4s), O (2s,2p), Fe (3d, 4s) electronic states were considered as valance. A 1×1×2 supercell of 60 atoms has been used to perform all the calculations of doped as well as pristine $YMnO_3$. The plane-wave cut-off energy of 680 eV and the convergence criteria for force tolerance and energy difference were set to 0.02 eV/A° and $10^{-8}$ eV, respectively. For meta-GGA calculations, energy convergence criteria were reduced accordingly to $10^{-4}$ eV. The Conjugate gradient approximation is utilized to optimize the lattice into its minimum energy state. During the optimization process, atomic positions, cell shape, and cell volume were allowed to vary. The calculations were carried out using a Gaussian smearing and a smearing width of 0.05 across the board. For static calculation using the PBE functional, the blocked-Davidson iteration scheme was set to ensure electronic convergence. The meta-GGA calculations were started from previously converged (PBE) wavefunctions, and a robust mixture of blocked-Davidson and RMM-DIIS algorithms was used. To ensure stable convergence, Kreker mixing was employed, and mixing parameters were carefully adjusted[41]. Important elastic and optical properties were obtained using VASPKIT[42]. The thermo-electric properties were computed using the semi-classical Boltzmann theory as implemented in the BoltzTrap2 code[43]. The magnetic and crystal structures of the materials were visualized using VESTA software[44].

## Results and Discussion

### Structural Properties:

The material under investigation in this work is hexagonal $YMnO_3$, which crystallizes in theP6$_3$cm space group, as illustrated in **Figure 1**. Four possible magnetic structures have been investigated: i) NM (Non-magnetic) as presented in **Figure 2**(a), in which non-spin polarized calculations have been employed; ii) FM (Ferro-magnetic), in which all Mn ions' spins are oriented along the c-direction as represented in **Figure 2**(b); iii) A-AFM (A-type Antiferromagnetic), in which Mn ions have ferromagnetic spins along the a-b plane but anti-ferromagnetic spins along the c-direction as represented in **Figure 2**(c); iv) G-AFM (G-type Anti-ferromagnetic), in which a frustrated collinear anti-ferromagnetic structure has been constructed by placing two up-spin Mn ions and one down-spin Mn ion in the basal plane as indicated in **Figure 2**(d). Although a triangular non-collinear magnetic structure is expected from the experiment[45] and previous theoretical considerations[46], in several studies, the magnetic structure of $YMnO_3$ has been approximated by a collinear frustrated AFM structure, as it is a reasonable representation of the actual material[47,48]. To determine the least energy structure, structural optimization is conducted and fitted with the third–order Birch–Murnaghan equation of state (EOS)[49]. Analyzing the total energy for different magnetic configurations, it is observed in **Figure 3** that the G-AFM configuration has the lowest energy among all four magnetic configurations studied. It is clear from Figure 3 that there is a significant difference in energy between the magnetic and non-magnetic configurations, and among all the studied configurations, the G-AFM is energetically most favorable, being 0.65 eV lower in energy compared to A-AFM. Additionally, the equilibrium cell volumes are 724.38 Å$^3$, 769.17 Å$^3$, 759.59 Å$^3$, and 752.41 Å$^3$ for Non-magnetic, FM, A-

AFM, and G-AFM, respectively. It is observed that lattice volume changes when the magnetic configuration is changed, which is an indication of the magneto-elastic response of the materials. From FM to G-AFM, a gradual decrease in volume is seen, which is consistent with a previous study conducted by Spladin et al.[50]. We have used the G-AFM magnetic configuration for doping and to calculate further properties of the materials.

The Wyckoff positions of the G-AFM configuration are shown in **Table 1**, along with an available experimental study[51]. These atomic arrangements control the tilting of the $MnO_3$ bipyramids and the shifts of Y ions, which play a key role in producing improper ferroelectricity and stabilizing the G-type antiferromagnetic ordering[17,52].

For doping of $YMnO_3$, the initial guess of the magnetic structure was taken to be G-AFM, and later on, it was allowed to relax to a favorable magnetic state. The estimated lattice parameters and volume of the G-AFM $YMnO_3$ and Fe-doped $YMnO_3$ are listed in **Table 2**. According to our calculation, the G-AFM state of $YMnO_3$ optimizes to a hexagonal $P6_3cm$ structure with slight distortion, possibly due to magneto-elastic response. The lattice structure of hexagonal $YFe_xMn_{1-x}O_3$ remains unchanged, although slight distortions are evident. In an experimental study conducted by Namdeo et al.[33], intra-lattice changes due to Fe substitution were reported, which is consistent with the distortion observed in our study. Additionally, an expansion of cell volume was followed by a rising concentration of Fe. As the doping level increases from $x = 0$ to $x = 0.25$, in-plane lattice constants a and b are decreased, while the out-of-plane lattice constant c increases (**Table 2**). This anisotropic behavior causes the c/a ratio to gradually rise, which is indicative of elongation along the c direction.

**Table 1** Wyckoff notations and fractional atomic coordinates for hexagonal $YMnO_3$.

| Atom | Wyckoff | Experimental[51] | | | G-AFM+GGA | | |
|---|---|---|---|---|---|---|---|
| | | x | Y | Z | x | y | z |
| Y(1) | 2(a) | 0.000 | 0.000 | 0.277 | 0.000 | 0.000 | 0.278 |
| Y(2) | 4(b) | 0.333 | 0.666 | 0.231 | 0.333 | 0.666 | 0.231 |
| Mn | 6(c) | 0.342 | 0.000 | 0.000 | 0.334 | 0.000 | 0.000 |
| O(1) | 6(c) | 0.301 | 0.000 | 0.161 | 0.306 | 0.000 | 0.164 |
| O(2) | 6(c) | 0.639 | 0.000 | 0.334 | 0.640 | 0.000 | 0.338 |
| O(3) | 2(a) | 0.000 | 0.000 | 0.480 | 0.001 | 0.001 | 0.478 |
| O(4) | 4(b) | 0.333 | 0.666 | 0.019 | 0.334 | 0.666 | 0.020 |

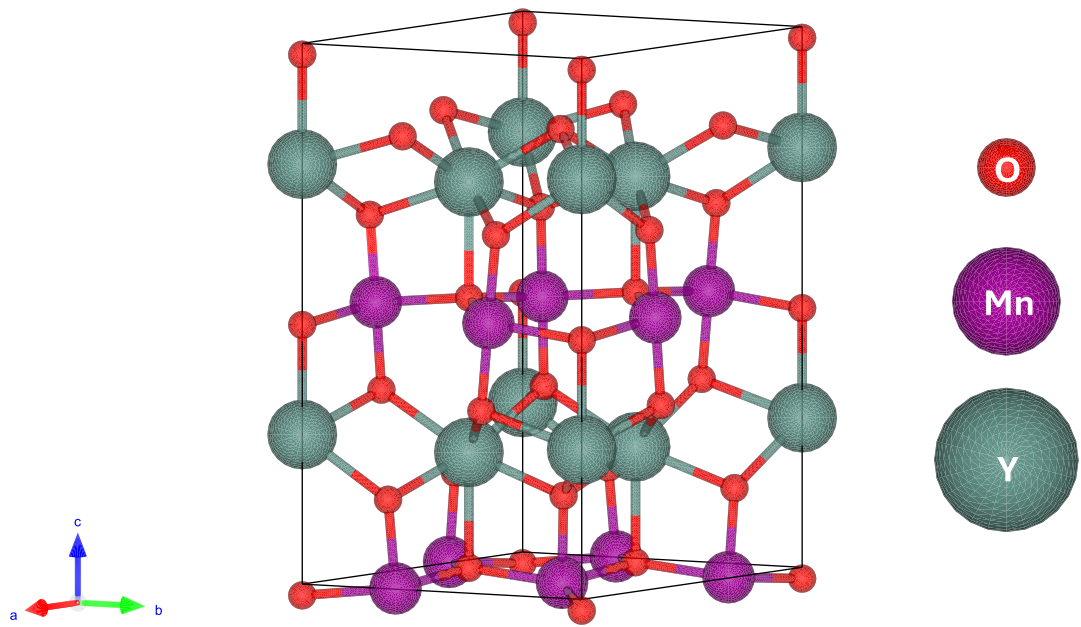

**Figure 1** Crystal structure of hexagonal YMnO$_3$ (Unit cell).

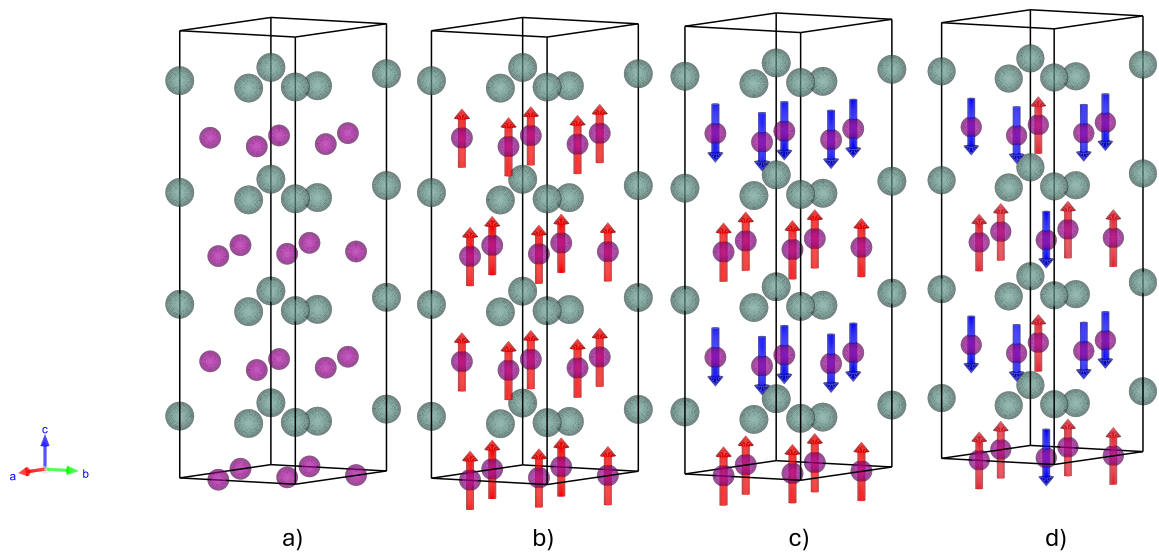

**Figure 2** Magnetic states of YMnO$_3$ (1×1×2 supercell): a) NM, b) FM, c) A-AFM, d) G-AFM (oxygen atoms are removed for better visibility).

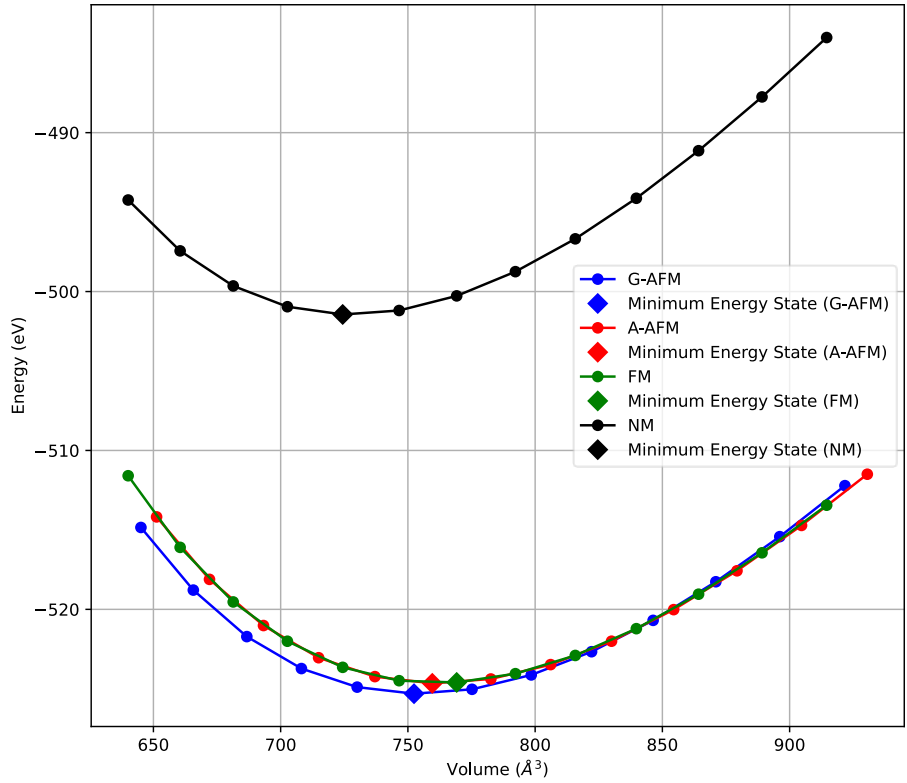

**Figure 3** Equation of States for different magnetic structures of YMnO$_3$.

**Table 2** Optimized structural parameters of YFe$_x$Mn$_{1-x}$O$_3$, namely lattice parameters (a, b, c) in Å, c/a ratio, and Volume Å$^3$.

| Concentration of YFe$_x$Mn$_{1-x}$O$_3$ | a (Å) | b (Å) | c (Å) | c/a | Volume (Å$^3$) |
|---|---|---|---|---|---|
| x = 0 | 6.143 | 6.143 | 11.510 | 1.874 | 376.155 |
| x = 0.08 | 6.138 | 6.138 | 11.532 | 1.879 | 376.329 |
| x = 0.16 | 6.134 | 6.134 | 11.551 | 1.883 | 376.493 |
| x = 0.25 | 6.129 | 6.129 | 11.575 | 1.889 | 376.702 |

## Electronic and Magnetic Properties:

We have used the mBJ potential to calculate spin-polarized electronic and magnetic properties, as the GGA functional underestimates the band gap for YMnO$_3$. The mBJ potential is employed only to calculate density of states, band structure, and optical properties, as it is not appropriate for the calculation of Hellmann-Feynman forces[39,40]. It has been reported in the literature that the band gap of YMnO$_3$ possesses values in the range between 1.4 eV ~ 2.1 eV. For instance, an optical band gap of 1.43 eV was reported by López-Alvarez et al.[53], and a band gap of 1.74 eV was reported by Inchara et al.[54], and 2.1 eV[29] by Polat et al[29]. The band gap of pure $YMnO_3$ in various magnetic ordering states is compiled in **Table 3**, along with the available experimental studies. It is confirmed from **Table 3** that our obtained band gap is in the range of experimentally measured values.

**Table 3** Band gap comparison of pristine $YMnO_3$ with experimental study.

| This study | | | | | Experimental |
|---|---|---|---|---|---|
| **NM (PBE)** | **FM (PBE)** | **A-AFM (PBE)** | **G-AFM (PBE)** | **G-AFM (mBJ)** | 1.4~2.1[29,53,54] |
| Metallic | Metallic | metallic | 0.59 | 1.88 | |

In **Figure 4**, band structures along with the density of states are displayed, where contributions from each element are represented via different colors for $YMnO_3$ and $YFe_xMn_{1-x}O_3$ (for x=0, 0.08, 0.17, and 0.25). The energy range that has been considered for the analysis of electronic band structure is -6 eV to 6 eV, and the energy has been shifted to accommodate the Fermi energy at 0 eV. In our study, $YMnO_3$ exhibits a comparable band structure to a study conducted by Chadli et al.[55] for both spin channels, as presented in **Figure 4**(a) (down spin) and **Figure 4**(b) (up spin). There is a direct band gap of 1.88 eV for both up and down spin channels, and the valence band maximum (VBM) is at -0.248 eV, and the conduction band minimum (CBM) is located at 1.63 eV. Moreover, this material shows antiferromagnetism. The nature of the band gap changes from direct to indirect upon Fe doping in the Mn sites, as is clear from **Figure 4**. Moreover, an anti-ferromagnetic to ferromagnetic transition is observed in the doped $YMnO_3$, possibly due to suppression of magnetic frustration. This result is consistent with the experimental result obtained by Ge et al.[56]. For the doping concentration of x = 0.08, the net magnetisation is 7.5 $\mu_B$/unitcell, and there is a significant difference in the band gap between spin up and down channels as depicted in **Figure 4**(c) and **Figure 4**(d). Spin-down electrons of $YFe_{0.08}Mn_{0.92}O_3$ have a valence band edge at -0.253 eV and conduction band edge at 1.349 eV, leading to an indirect bandgap of 1.60 eV. Likewise, for spin-up electrons, VBM and CBM are at -0.624 eV and 1.902 eV, respectively, leading to an indirect band gap of 2.53 eV. For the 17% Fe-doped $YMnO_3$, spin-resolved structures are presented in **Figure 4**(e) (spin down) and **Figure 4**(f) (spin up), respectively; the net magnetisation in this structure is 1 $\mu_B$/unitcell. The values of the band gaps are 1.58 eV (indirect) and 1.88 eV (direct) for the spin-down and spin-up bands, respectively. The corresponding band edges of spin-down and spin-up channels are at -0.250 eV (VBM), 1.633 eV (CBM), and -0.248eV (VBM), 1.339 eV (CBM), respectively. For the 25% Fe-doped structure, $YFe_{0.25}Mn_{0.75}O_3$, the net magnetisation is found to be 8.5 $\mu_B$/unitcell, the spin-down bands exhibit a band gap of 1.19 eV (**Figure 4**(g)) due to their VBM appearing at -0.243 eV and CBM appearing at 0.951 eV. Similarly, spin-up bands have the top of the valence band located at -0.809 eV and the bottom of the conduction band located at 0.476 eV, which gives a band gap of 1.29 eV (indirect) as seen in **Figure 4**(h).

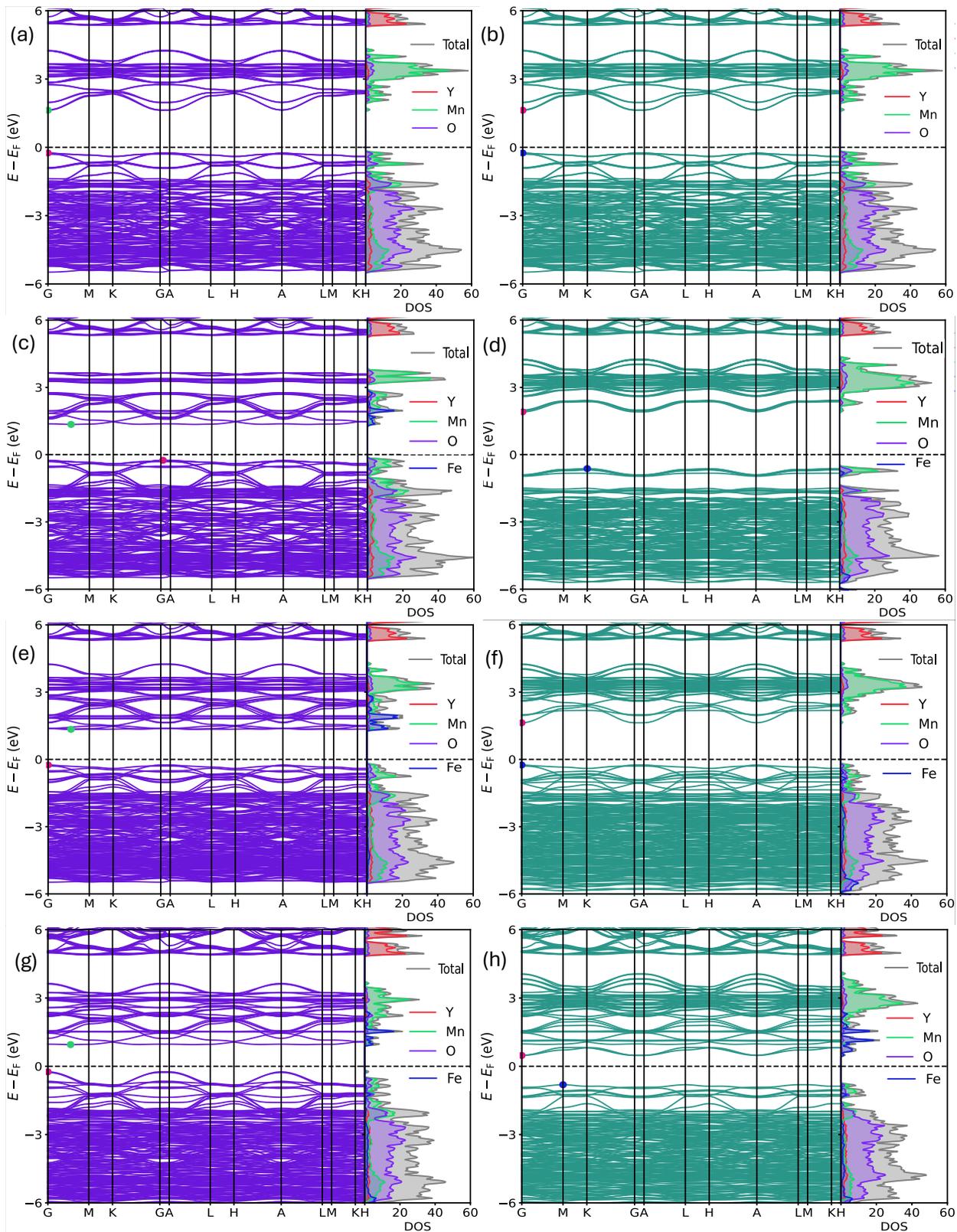

**Figure 4** Band structure and density of states for YFexMn1-xO3 for spin down channel where (a) x = 0, (c) x = 0.08, (d) x = 0.17, and (e) x = 0.25, likewise for spin up channel (b) x = 0, (d) x = 0.08, (e) x = 0.17, and (f) x =0.25. (VBM and CBM are represented via colored points)

Overall, a decreasing trend in the band gap is observed with increasing Fe concentration in the YMnO$_3$ for the spin-down channel, as presented in **Figure 5**. A similar trend was observed for Os doping in YMnO$_3$[29]. On the other hand, for the spin-up channel, there is an initial increase in the band gap followed by a declining trend (**Figure 5**). The initial increase in band gap can be attributed to the rise of net magnetization, which in turn introduces asymmetry between the band gaps of two spin orientations[57]. A similar non-monotonic change of band gap with respect to doping concentration is observed in CaTiO$_3$ in a study conducted by Deluque-Toro et al.[58]. Additionally, for the injection of Fe into YMnO$_3$, flat bands near the Fermi energy are introduced as seen in **Figure 4**(c), **4**(e), and **4**(g). The presence of flat bands near the Fermi level reduces carrier mobility due to large effective mass carriers. It also presents opportunities for interesting quantum phases due to the localisation of electrons[59].

The density of states and projected density of states for YFe$_x$Mn$_{1-x}$O$_3$ (where x =0, 0.08, 0.17, and 0.25) are analysed within a range of -6 eV to 8.5 eV to better understand the orbital contribution to electronic properties. From **Figure 6**(a), it is visible that the 2p orbital of O mostly populates the conduction band of YMnO$_3$, Y_4d states dominate in the higher range of the conduction band, and a significant contribution of Mn_3d states is seen near the Fermi level, which is the primary reason for choosing B site for doping so that the band gap of YMnO$_3$ can be tuned. As band edges are formed mostly due to contributions from Mn 3d and O 2p states, substituting a different cation in the B site can alter these states, changing the alignment of VBM/CBM. Additionally, equal contribution from the up and down spin below the Fermi level, as seen from **Figure 5**(a), ensures the antiferromagnetic nature of YMnO$_3$. For the doping concentration of x = 0.08 (**Figure 5**(b)), the contribution of Y_4d states in the deeper conduction band and O_2p states in the valence band is observed. However, near the Fermi level, a small contribution of Fe_3d states is observed in addition to Mn_3d states. Projected density of states for YFe$_{0.17}$Mn$_{0.83}$O$_3$ is presented in **Figure 5**(c), where spin-down and spin-up channel has unequal contributions, resulting in a ferromagnetic nature, and the contribution of Fe_3d states is greater compared to the x=0 and x=0.08 cases. For x = 0.25, a more pronounced contribution of Fe_3d states has been observed near the Fermi level, along with decreased contribution from Mn_3d states as seen from **Figure 5**(d). Furthermore, an asymmetric nature is observed between spin-up and spin-down states.

The compound's net magnetic moment and the atom's projected magnetic moment are calculated using the meta-GGA mBJ potential. **Table 4** depicts total magnetic moments and average magnetic moments for different elements in YFe$_x$Mn$_{1-x}$O$_3$ (for x =0, 0.08, 0.17, and 0.25).

The average value of the projected magnetic moment for the Mn ion is 3.65 µ$_B$, as represented in **Table 4**, which is in good agreement with the available experimental data (2.9~3 µ$_B$[60]). The magnetic moment of the Mn ion remains more or less the same in pure or Fe-doped structures. Fe ions have magnetic moments within a range of 3.76~4.08 µ$_B$. YMnO$_3$ exhibits a ferromagnetic nature with the inclusion of Fe ions.

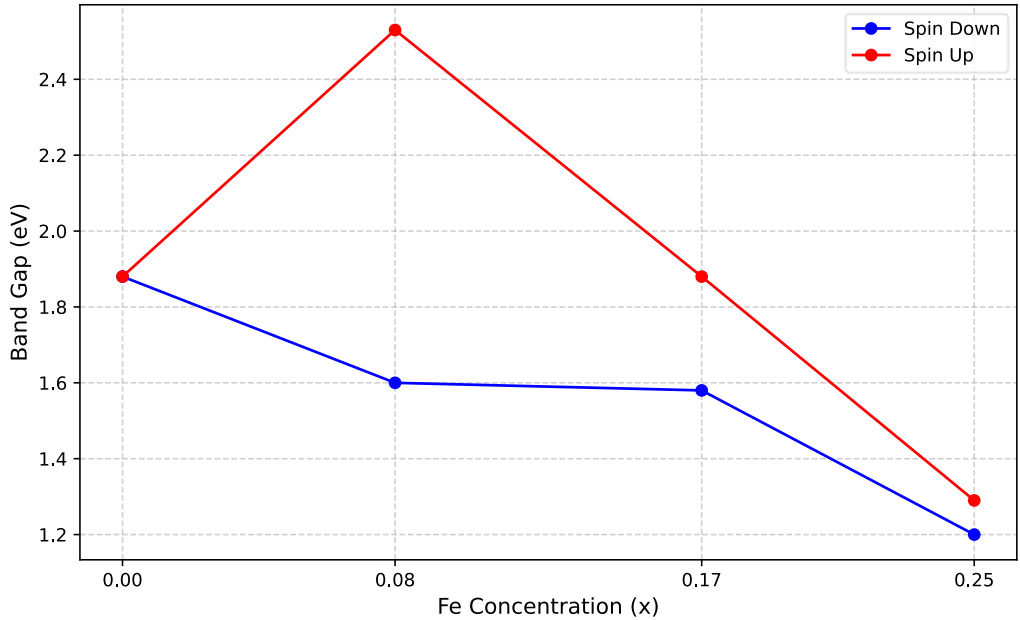

**Figure 5** Band-Gap vs Fe concentration for spin-up and spin-down channels.

**Table 4** Projected magnetic moment and Total magnetic moment of YFe$_x$Mn$_{1-x}$O$_3$.

| Concentration | x = 0 | x = 0.08 | x = 0.17 | x = 0.25 |
|---|---|---|---|---|
| µ$_Y$ | 0.00 | 0.01 | 0.00 | 0.00 |
| µ$_{Mn}$ | 3.65 | 3.65 | 3.49 | 3.66 |
| µ$_{Fe}$ | - | 4.03 | 3.76 | 4.08 |
| µ$_O$ | 0.03 | 0.08 | 0.04 | 0.08 |

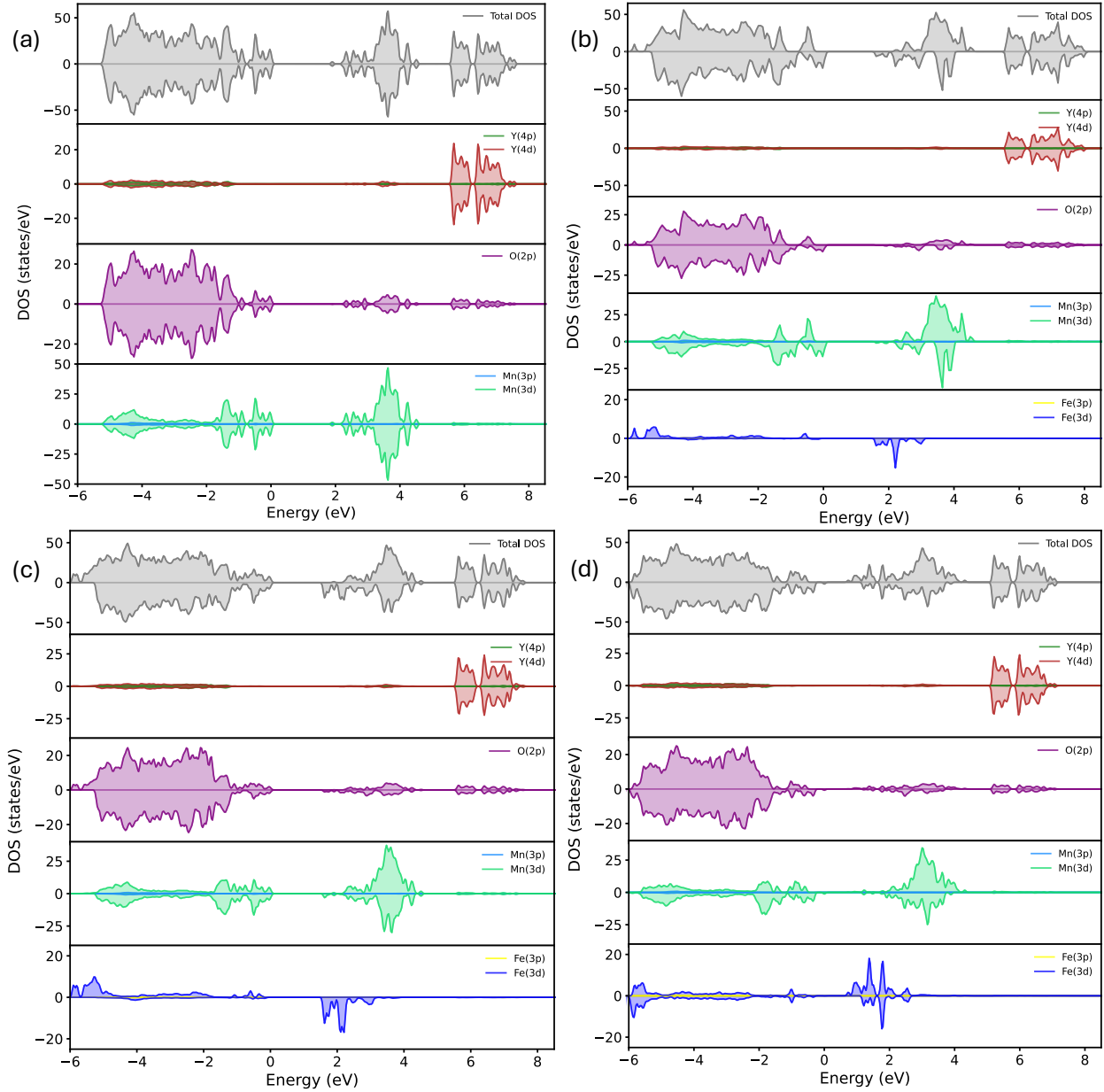

**Figure 6** Density of states and Projected density of states of YFe$_x$Mn$_{1-x}$O$_3$, where (a) x = 0, (b) x = 0.08, (c) x = 0.17, (d) x = 0.25.

## Optical Properties:

Optical Properties of a material are essential to understand its utilization in optoelectronics and photovoltaic applications. Therefore, the frequency-dependent response of various optical features of YFe$_x$Mn$_{1-x}$O$_3$, such as dielectric function, refractive index, reflectivity, absorption, photoconductivity, loss function, and extinction coefficients, has been studied extensively in the range of photon energies of 0-30 eV.

The frequency-dependent dielectric function of a material, $\varepsilon(\omega) = \varepsilon_1(\omega) + i\varepsilon_2(\omega)$, has a major impact on its applications[61]. Here, the dispersion, or how the speed of light changes in a material, can be described by using the real part $\varepsilon_1(\omega)$, and optical absorption can be described by using the imaginary part $\varepsilon_2(\omega)$. As hexagonal crystal structures possess anisotropy in the dielectric tensor[62], optical properties are analyzed along the a-b (i.e., xx component) and c (i.e., zz component) directions separately as presented in **Figure 7**(a) and **7**(b), respectively. The static dielectric constant $\varepsilon(0)$ has a value of 2.78 along the a-b direction, which is lower than the experimental value of 4.5[63]. It is observed from **Figure 7**(a) that $\varepsilon(0)$ increases with the increase of Fe concentration, the highest obtained value of $\varepsilon(0)$ = 3.24 corresponding to a Fe concentration of 0.25. The recombination rate of charge carriers is highly dependent on the static electric constant; a suppression of charge carrier recombination rate is observed for materials with a higher dielectric constant, which impacts the overall efficiency of devices like solar cells, photodetectors, and light-emitting diodes[64]. Hence, anticipation can be made that Fe doping in $YMnO_3$ can have enhanced optoelectronic properties than the pure $YMnO_3$. The static dielectric constant $\varepsilon(0)$ along the c-direction is 3.35 for pristine $YMnO_3$ as presented in **Figure 7**(b), being in good agreement with the experimental value of 4.8[63] and the doped materials have a value of $\varepsilon(0)$ between 3.47 ~3.55. As depicted in **Figure 7**(b), the real part of the dielectric function does not show significant variation for different doping concentrations in the c-direction. The imaginary part of the dielectric function for $YMnO_3$ along the a-b direction has a significantly large value for the photon energy of 1.5 to 10 eV, as shown in **Figure 7**(c), which indicates a good absorption in the visible and UV range. Additionally, the first peak of $\varepsilon_2$ is observed at 1.9 eV, which aligns well with experimental observation of a peak at 1.6 eV[63]. This first peak shifts towards 2 eV for Fe-doped $YMnO_3$ with an increasing amplitude with increased concentration of Fe, and 25% Fe concentration portrays the highest peak. For $\varepsilon_2$ along the c direction, we can observe a peak at 6 eV for pure $YMnO_3$ as presented in **Figure 7**(d) that corresponds to the experimentally observed peak at 5.2 eV[63]. A smaller peak observed experimentally at 1.7eV is often absent in computational and theoretical studies[65,66]. It is also visible that this peak decreases in amplitude with increasing concentration of Fe in doped $YMnO_3$.

Optical properties such as absorption coefficient α(ω), extinction coefficient K(ω), reflectivity R(ω), refractive index η(ω), energy loss spectra L(ω), and optical conductivity σ(ω) can be obtained from the dielectric function $\varepsilon(\omega)$[67,68].

The absorption coefficient α(ω) gives information about how well these materials capture photons of various energies and how they absorb incoming radiation by moving electrons from lower to upper energy levels[69]. **Figure 7**(e) shows the absorption coefficient α(ω) for $YFe_xMn_{1-x}O_3$ along the a-b direction, where significant absorption is observed between 1.5 eV to 13 eV. Pure YMnO$_3$ shows a maximum peak value of 8.8×10$^5$ at 9.9 eV, whereas Fe-doped YMnO$_3$ shows the highest peak value of 1.14×10$^6$ at 10.5 eV, corresponding to concentration x = 0.08. As it is seen from **Figure 7**(e), there are several peaks in the range 20-30 eV, which indicate transitions from deeper valence states to conduction states[70]. Due to the presence of flat bands in the deep conduction band (**Figure 4**), these transitions are possible when incoming photons have sufficient energy[62]. Similarly, absorption along the c-direction shows significant absorption of all the materials in the UV range (3-13 eV), with the highest absorption peak for Fe concentration of x = 0.08 in **Figure 7**(f). The extinction coefficient K(ω) for various concentrations of Fe doping is shown in **Figure 7**(g) (along the a-b direction), which determines the strength of the material to absorb or scatter photons and transmit photons through the material. It shows peaks at 1.9 eV and 2 eV, respectively, for undoped and doped YMnO$_3$, consistent with the peaks observed in $\varepsilon_2$ (**Figure 7**(b)). The highest peak is observed for Fe concentration of x = 0.08 at 10.5 eV. Several peaks in the UV range demonstrate the deep penetration and diffusion of UV light for the compounds. Because of their potent optical absorption in the ultraviolet region, the materials show significant promise for photocatalysis applications[71]. **Figure 7**(h) presents the variation of K(ω) along the c-direction. Here, undoped YMnO$_3$ shows the highest extinction among all the materials, and it is observed at 6.15 eV.

Another essential optical property of a material is reflectivity R(ω), which characterizes how the surface of a material responds to incident electromagnetic waves and is crucial for characterizing materials and assessing their possible applications[72]. It is the ratio of the amount of light energy incident on a material's surface to the amount of light energy reflected from that surface. The materials' high reflectivity suggests that they are suitable for applications involving shielding and anti-reflective coatings. **Figure 8**(a) shows the variation in reflectivity R(ω) with photon energy along the a-b-direction for YFe$_x$Mn$_{1-x}$O$_3$. The reflectivity for zero energy R(0) is 0.063, 0.075, 0.078, and 0.082 for Fe concentration of x = 0, 0.08, 0.17, and 0.25, respectively. These compounds show maximum reflectivity R(ω) between the energy range of 6 eV to 13 eV. As illustrated in **Figure 8**(b), the reflectivity of YFe$_x$Mn$_{1-x}$O$_3$ along the c-direction varies with incident photon energy.

Reflectivity at zero energy is 0.086 for undoped YMnO$_3$ and 0.094, 0.091, 0.093 for doping concentrations of x = 0.08, 0.17, and 0.25, respectively.

All the compounds demonstrate the highest reflectivity within the energy range of 4 to 13 eV. The efficacy of a material in optical applications is significantly influenced by its refractive index. For example, materials with high refractive indices are perfect for photovoltaic cells[73]. When designing optical components like thin films, pigments, and lenses, the refractive index η(ω) is directly related to microscopic atomic interactions[74].

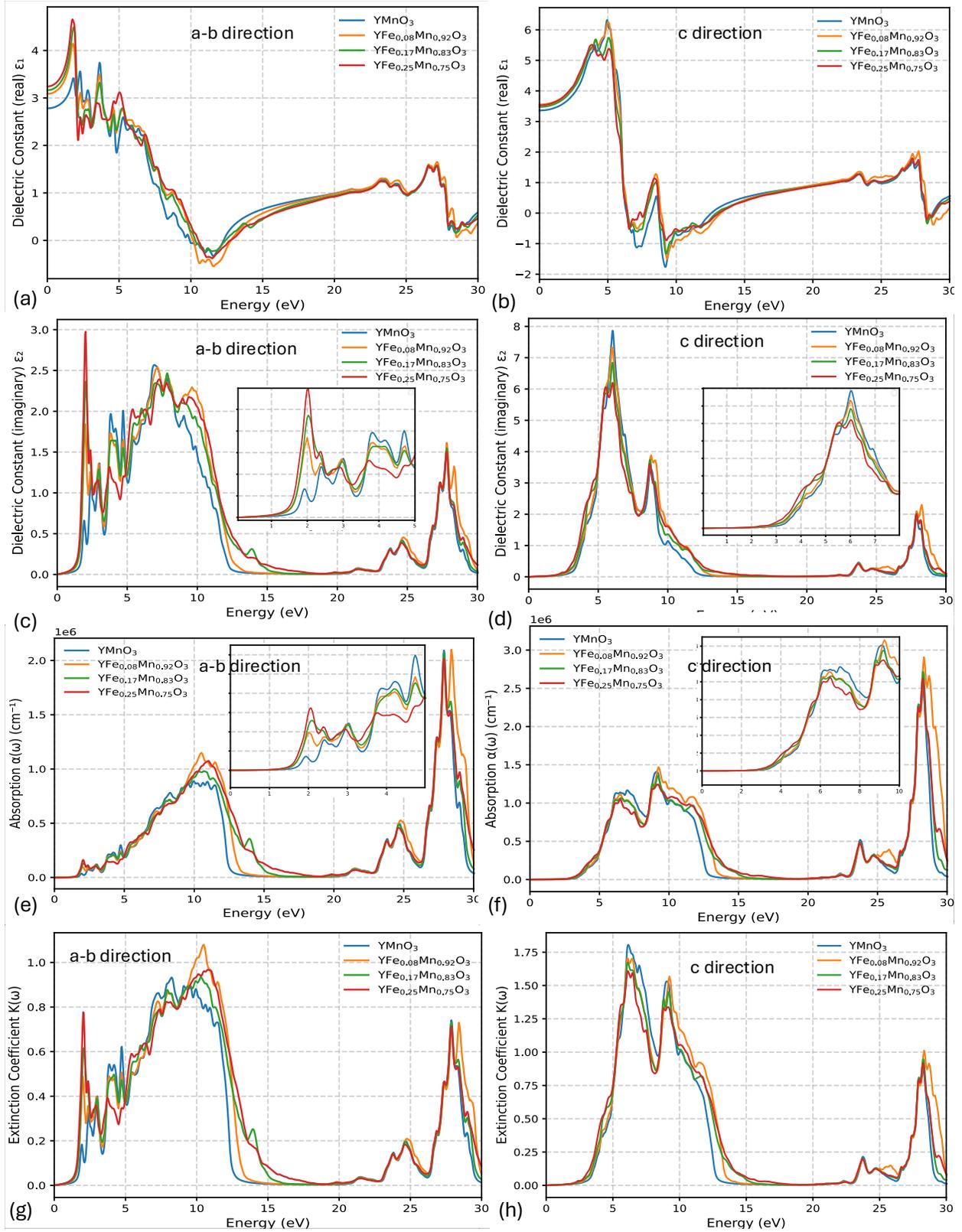

**Figure 7** Optical properties of YFe$_x$Mn$_{1-x}$O$_3$ along the a-b direction (a) real part of the dielectric function, (c) imaginary part of the dielectric function, (e) absorption spectrum, (g) extinction coefficient, and along the c

direction (b) real part of the dielectric function, (d)imaginary part of the dielectric function,( f) absorption spectrum, (h) extinction coefficients.

The refractive index η(ω) for doped and undoped $YMnO_3$ along the a-b-direction is presented in **Figure 8**(c). The refractive index at zero energy, η(0), for different concentrations is shown in **Table 5**. The material shows a peak around 1.9 eV, after that the value drops to the lowest around 12~13 eV for every doping concentration, so the studied materials are less transparent for higher photon energy. Similarly, the refractive index η(ω) along the c-direction is presented in **Figure 8**(d). The compounds reach a peak value of refractive index at around 5 eV and consequently reach the lowest point around 12~13 eV. The refractive index η(ω) shows close resemblance to the dielectric constant $\varepsilon_1(\omega)$, as expected (**Figure 7**(a & b)).

An electron loses energy as it passes through a substance by interband transitions, plasmon and phonon excitations, and inner shell ionization[75]. The electrons in the material can move quickly enough to produce plasma oscillations that, in effect, screen out the external electric field when the plasma frequency is higher than the incident electromagnetic wave's frequency[76]. Multiple peaks in the UV region of the L(ω) spectra (along the a-b direction) are shown in **Figure 8**(e), suggesting that, compared to the visible region, this region experiences significantly higher electron losses. A similar trend is seen for the loss spectrum along the c-direction as depicted in **Figure 8**(f). When the right frequency of light strikes a material's surface, electrical conduction starts, producing optical conductivity σ(ω)[75,77]. Optical conductivity σ(ω) for $YFe_xMn_{1-x}O_3$ along the a-b direction and c direction are illustrated in **Figure 8**(g) and **Figure 8**(h), respectively. In both cases, significant conductivity is seen within the energy range of 4 to 10 eV and 27 to 29 eV of photon energies.

**Table 5** Optical parameters of $YFe_xMn_{1-x}O_3$ corresponding to zero energy.

| concentration | x = 0 | | x = 0.08 | | x = 0.17 | | x = 0.25 | |
|---|---|---|---|---|---|---|---|---|
| | ab | c | ab | c | ab | c | ab | C |
| $\varepsilon_1(0)$ | 2.78 | 3.35 | 3.09 | 3.56 | 3.16 | 3.48 | 3.24 | 3.53 |
| η(0) | 1.67 | 1.83 | 1.75 | 1.88 | 1.78 | 1.86 | 1.80 | 1.87 |
| R(0) | 0.06 | 0.08 | 0.07 | 0.09 | 0.08 | 0.09 | 0.08 | 0.09 |

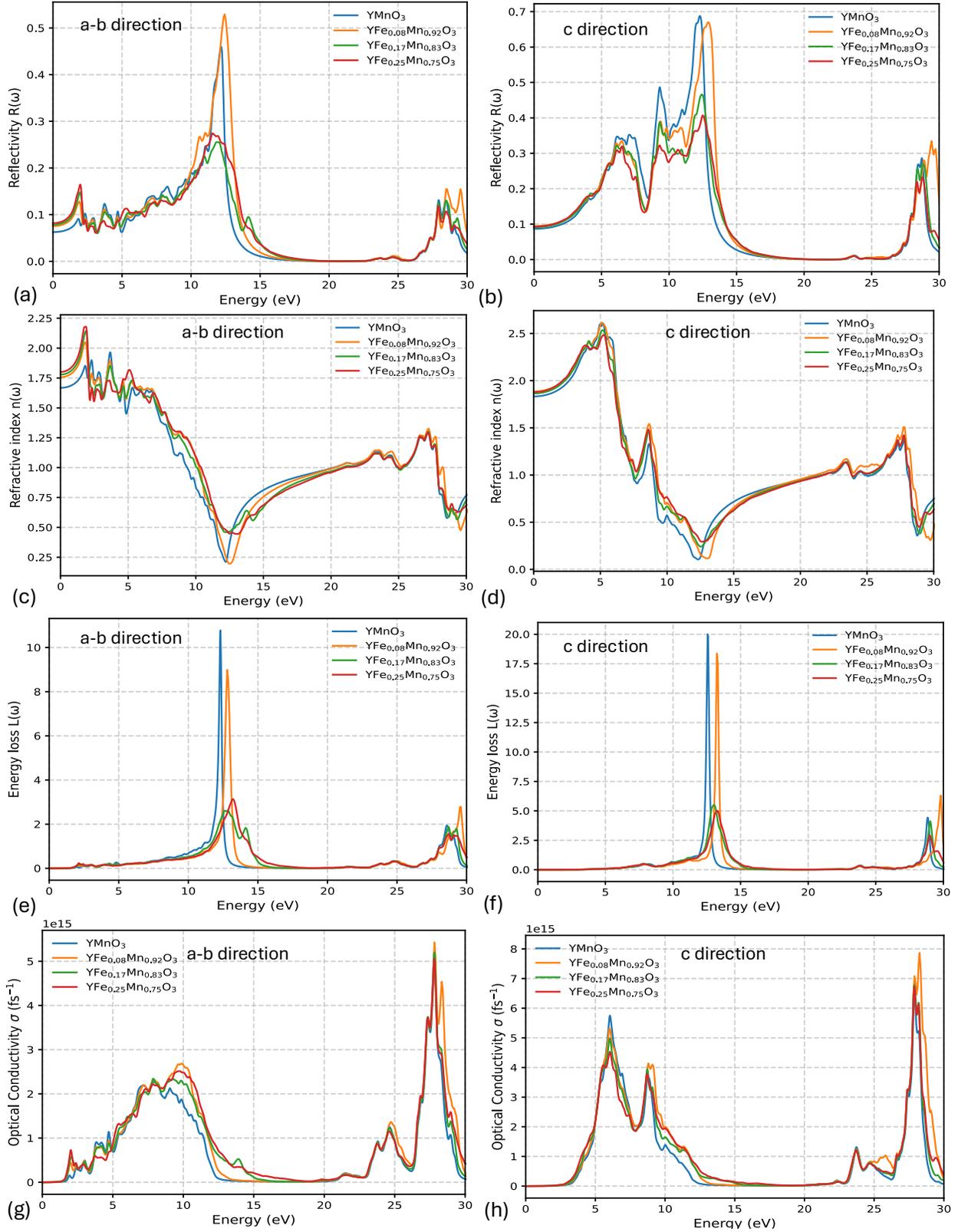

**Figure 8** Optical properties of YFe$_x$Mn$_{1-x}$O$_3$ along the a-b direction (a) Reflectivity, (c) Refractive index, (e) Energy loss, (g) Optical conductivity, and along the c direction (b) Reflectivity, (d) Refractive index, (f) Energy loss, (h) Optical conductivity.

## Thermoelectric properties:

Converting waste heat into useful energy requires an understanding of the thermoelectric properties of materials. To address the global energy shortages, reduce the harmful emissions, and convert excess heat into electricity, thermoelectric compounds present a promising solution[78,79]. Seebeck coefficient (S), Specific Heat ($C_v$), Magnetic Susceptibility ($\chi$), Electrical Conductivity ($\sigma/\tau_0$), Thermal conductivity ($\kappa/\tau_0$), Hall coefficient ($R_h$), and Power Factor (PF) are calculated using the constant relaxation time approximation as implemented in the Boltztrap2 code[43] for a temperature range of 300K to 1000K.

The Seebeck coefficient (S) is essential for figuring out a material's thermoelectric efficiency, which describes its ability to generate electrical potential. It measures the voltage that a material can generate when it is subjected to temperature gradients across it. The total Seebeck coefficient of all the studied materials is illustrated in **Figure 9**(a). Both undoped and Fe-doped $YMnO_3$ exhibit an increase in the Seebeck coefficient with increasing temperature, as would be expected for a semiconductor[80]. Additionally, as seen in **Figure 9**(a), the doped system has an overall smaller value of the Seebeck coefficient than the undoped one, as doping introduces additional charge carriers in a material, which in turn reduces the Seebeck effect[81]. In **Figure 9**(b), the conductivity per relaxation time ($\sigma/\tau_0$) is presented for all the doping concentrations; it is observed from **Figure 9**(b) that the conductivity increases with respect to temperature for all the materials. The increase in conductivity with temperature can be explained as the charge carriers gain energy due to the temperature rise to overcome the bandgap to reach the conduction band. This trait affirms the materials' semiconducting nature. Furthermore, due to the additional charge carrier introduced because of doping, the conductivity has increased significantly in the doped systems as compared to the undoped system, except for the Fe concentration of x = 0.17. For 17% Fe-doped $YMnO_3$ conductivity has dropped slightly at room temperature, resulting from the presence of several flat bands near the Fermi-energy.

The Hall coefficient ($R_H$) is a basic property of a material that assists in identifying the nature of the charge carrier of a material. From **Figure 9**(c), it is visible that all the studied materials have a positive value of Hall coefficient at room temperature, confirming their p-type behavior. This is consistent with **Figure 4**, where the Fermi energies corresponding to the band structure of doping concentration x = 0, 0.08, and 0.17 are located close to their valence band. On the contrary, for x = 0.25, the Fermi energy is located closer to the valence band for spin down, but for the spin-up channel, it is closer to the conduction band, as seen in **Figure 4**(g) and **Figure 4**(h). However, the combined effect still favors p-type activity.

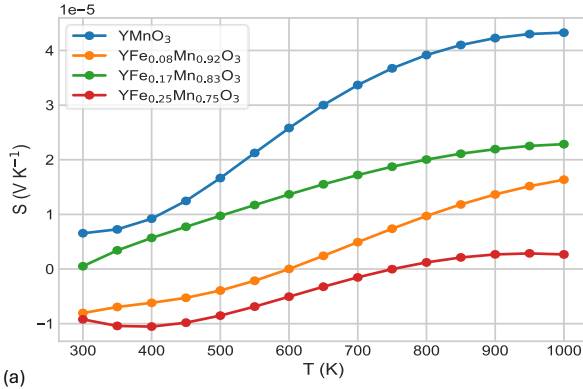
(a)

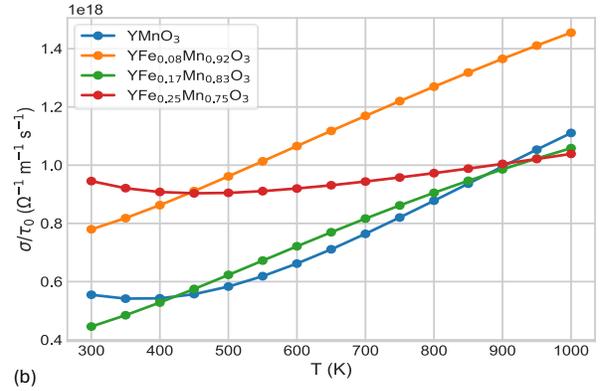
(b)

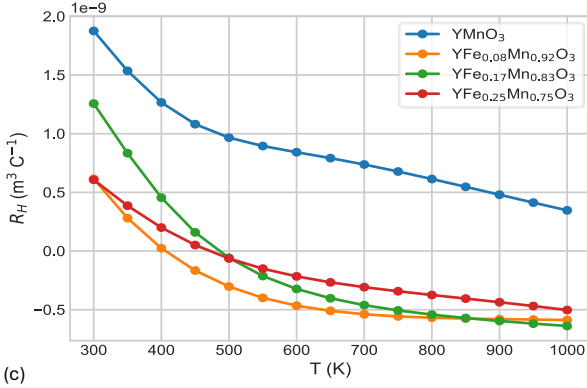
(c)

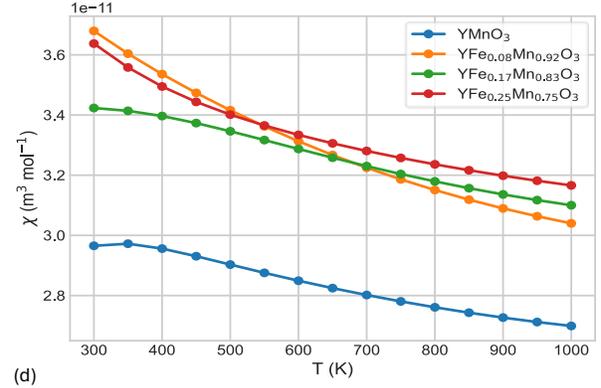
(d)

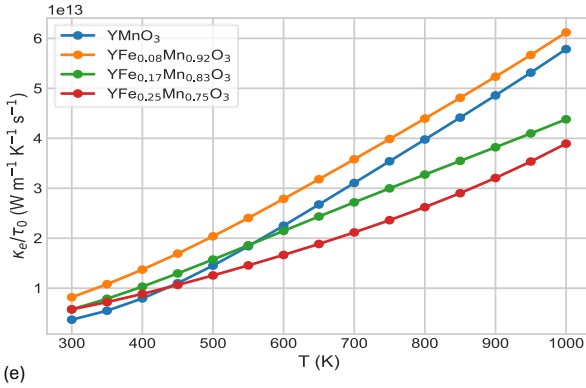
(e)

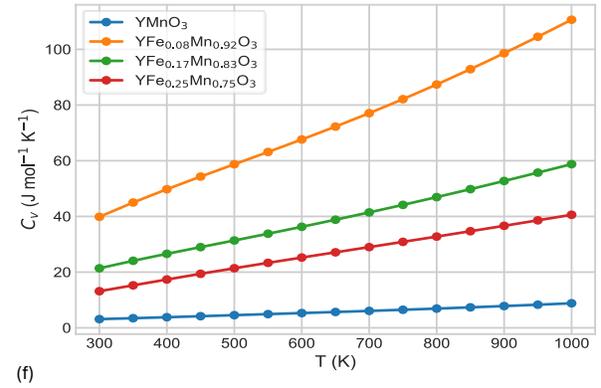
(f)

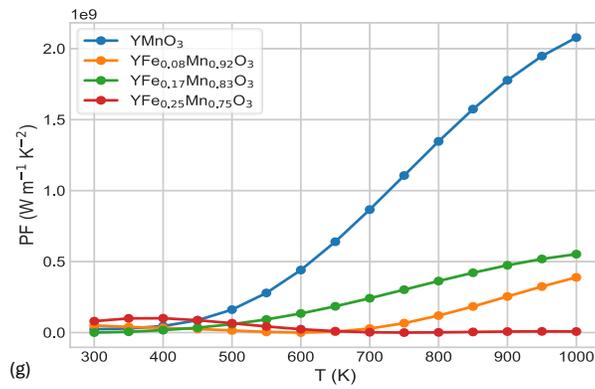
(g)

**Figure 9** Thermoelectric properties of YFe$_x$Mn$_{1-x}$O$_3$ with varying temperature a) Seebeck coefficient (S), b) Electrical Conductivity (σ/τ$_0$), c) Hall coefficient (R$_H$), d) Magnetic Susceptibility (χ), e) Thermal Conductivity (κ$_e$/τ$_0$), f) Specific heat (C$_v$), g) Power Factor (PF).

When subjected to a magnetic field, how much a material is prone to become magnetized is quantified by the magnetic susceptibility (χ). **Figure 9**(d) shows that the magnetic susceptibility of YFe$_x$Mn$_{1-x}$O3 decreases with increasing temperature for all materials. This is because, as with the rise of temperature, the magnetic domains become increasingly disordered, which reduces the net magnetization. This pattern follows the Curie-Weiss law of magnetization. It is also observed from **Figure 9**(d) that the magnetization has increased with the increase of Fe-doping concentration, as expected, because of weakened anti-ferromagnetism.

Another crucial factor in determining a material's thermoelectric efficiency is its electronic thermal conductivity per relaxation time (κ/τ$_0$), which gauges the material's capacity to transfer heat via electron transport. Total heat conductivity (κ) is the sum of electronic (κ$_e$) and phononic (κ$_{ph}$) heat conductivity. Regardless of its impact on the material's overall thermal conductivity, the phononic component (κ$_{ph}$) is ignored in the current study's calculation of thermal conductivity due to the limitations imposed by the Boltztrap2 code. From **Figure 9**(e), we can see that the thermal conductivity (κ/τ$_0$) increases with temperature. This increase is the result of increasing molecular vibrations due to the temperature rise. Additionally, the obtained κ/σ ratio of the materials is in the order of 10$^{-5}$, which is consistent with the Wiedemann-Franz law. When paired with a high Seebeck coefficient, this value suggests that the electronic contribution to thermal conductivity is similar to what is anticipated for good electrical conductors, which is advantageous for thermoelectric performance[81]. A material's specific heat (C$_v$), which indicates how much thermal energy it can store, is the quantity of energy needed to raise its temperature by one degree per unit mass. **Figure 9**(f) shows an increasing trend in specific heat for the materials, where YFe$_{0.08}$Mn$_{0.92}$O$_3$ has the highest specific heat. As the Dulong-Petit limit is not reached, specific heat is expected to rise with temperature.

Power factor (PF) has also been calculated, where PF = S$^2$σ. It measures a material's capacity to generate electrical power from a temperature gradient. As illustrated in **Figure 9**(g), pure YMnO$_3$ possesses the highest Power factor. Power factor rises initially for all the concentrations, then saturates after a point, or slowly decreases. This study shows that doping significantly increases conductivity, but that doped YMnO$_3$'s power factor has dropped as a result of a drop in the Seebeck coefficient.

## Conclusion

Fe-doped $YMnO_3$ has been studied by utilizing the first-principles approach, which demonstrates that doping Fe on the B-site of $YMnO_3$ introduces slight structural distortions. It is observed that Fe-doping reduces the band gap of $YMnO_3$ and tends to shift magnetization towards Ferromagnetic by reducing the magnetic frustration. The reduction of the electronic bandgap is observed for undoped materials (1.88 eV) to a value of 1.19 eV for a Fe concentration of x = 0.25. Additionally, doped compounds show improved absorption and dielectric characteristics. Fe doping raises the static dielectric constant, which lowers the devices' recombination losses. Absorption in the visible and UV range gains intensity due to doping. A reduced bandgap with robust absorption presents promising utilization in photovoltaic devices. Add to that all the materials possess high reflectivity, which is suitable for photovoltaic devices. Thermoelectric analysis reveals enhanced conductivity due to the injection of extra charge carriers because of doping. Although conductivity is improved, due to the decline of the Seebeck coefficient, an overall lower power factor for doped $YMnO_3$ is obtained compared to pristine $YMnO_3$.

Overall, multifunctional tuning of $YMnO_3$ is made possible by Fe doping. While the higher electrical conductivity indicates some utility in waste-heat conversion, even as the Seebeck coefficient is reduced, the calculated narrowing of the bandgap and enhanced $\varepsilon(0)$ support its use as a lead-free ferroelectric absorber in solar cells. All things considered, this thorough computational analysis demonstrates the tenability of various properties of $YFe_xMn_{1-x}O_3$ and offers design recommendations for its application as a thermoelectric material.

## Acknowledgement

The authors acknowledge the contribution of Computational Physics Lab, Department of physics, University of Dhaka, and Bangladesh Research and Education Network (BdREN) for providing computational facilities. K. M. Kamal also acknowledges the contribution of the Ministry of Science and Technology, Bangladesh for giving him fellowship while working on this project.